# Simultaneous polarization attraction and Raman amplification of a light beam in optical fibers

**Philippe Morin, Stéphane Pitois and Julien Fatome***

*Laboratoire Interdisciplinaire Carnot de Bourgogne (ICB)*

*UMR 6303 CNRS / Université de Bourgogne, 9 av. Alain Savary, 21078 Dijon, France*

*\*Corresponding author: jfatome@u-bourgogne.fr*

In this paper, we demonstrate that it is possible to combine both Raman amplification and polarization attraction of a signal wave in a single optical fiber by means of a counter-propagating scheme. Experiments were performed near 1550 nm in continuous wave regime and by means of a 10-Gbit/s return-to-zero signal injected in a 20 km-long low-PMD optical fiber. Complete repolarization and 6.7 dB amplification of the signal wave were achieved by injecting a 850-mW 1480-nm counter-propagating polarized pump wave.

## 1. Introduction

With the development of ultra-high-bit-rate and transparent optical communication systems and their associated complex modulation formats and signal processing issues, several research groups have focused their attention on the conception of novel devices allowing a fast and all-optical control of the state of polarization (SOP) of a light beam. Indeed, in order to overcome electronic bottleneck and to go beyond the limit of current electronic feedback systems [1]-[2], nonlinear effects taking place in optical fibers have received much attention as a possible way to develop ultrafast polarization processing. For instance, Martinelli *et al*. have reported in refs. [3]-[4] an original effect of polarization pulling in a fiber Raman amplifier based on polarization dependent gain of the amplification process [3]-[9] whereas Thevenaz *et al.* have demonstrated in ref. [10] that an all-optical polarization control can be obtained through stimulated Brillouin back-scattering occurring in optical fibers [10]-[16]. Recently, it has also been reported that an all-optical polarization stabilization of a telecom signal can be achieved thanks to a counter-propagating four-wave-mixing (FWM) process taking place in a low-birefringence, i.e. low polarization mode dispersion (PMD), optical fiber [17]-[20]. This nonlinear polarizer was experimentally observed by simultaneously injecting an arbitrary polarized signal wave and a polarized control pump beam via a counter-propagating scheme. More precisely, we have shown in these previous works that, thanks to the FWM process, the pump wave induces a unidirectional exchange of energy between the orthogonal polarization components of the signal wave, thus leading to the so-called polarization attraction or "funnel" effect [21]-[28]. Here, an important feature of this phenomenon is that the polarization attraction process is supposed to act even though the pump and signal waves have different frequencies [27]-[29], basically because the four-wave mixing effect remains constantly phase-matched whatever the frequency detuning between the two waves [27]-[29].

Based on this last property, the aim of this work is thus to provide a proof-of-principle demonstrating that it is possible to combine both polarization attraction and Raman amplification in a unique segment of optical fiber thanks to a polarized frequency-shifted counter-propagating optical pump wave. We would like to emphasize that the possibility of coupling this two fundamental optical functions in such a device has already been reported in a previous work, in which pioneering experiments were performed in the visible domain by means of kW power nanosecond pulses injected in a few meters of isotropic optical fiber [29]. In this novel work, we focus our attention on the possibility to achieve such a process in the C-band around 1.55 µm, with both continuous waves and telecommunication signals, thus demonstrating the full potential of this device for practical applications. The first part of the manuscript is devoted to theoretical considerations dealing with Raman amplification and polarization attraction process occurring in an optical fiber pumped by two counter-propagating optical waves. In next section, we describe the experimental set-up used to observe these two combine effects. Finally, third and fourth parts of the paper are devoted to experimental results obtained either with a continuous optical signal and a 10 Gbit/s Return-to-Zero (RZ) telecom signal, respectively.

## 2. Theoretical Considerations

Basically, the device presented in this paper and combining both Raman amplification and polarization stabilization, is based on the nonlinear interaction between two counter-propagating waves occurring in an isotropic or low PMD optical fiber (see Fig .1).

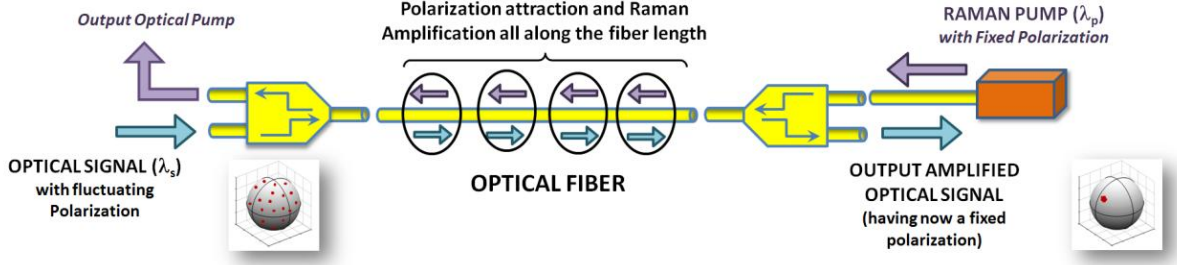

**Fig. 1:** Principle of the device allowing both polarization attraction and Raman amplification

For sake of simplicity, and without loss of generality, we will consider in this theoretical part the case of a perfectly isotropic fiber. To this aim, let us first consider two laser beams, a forward signal beam and a backward pump beam that counter-propagate with the same wavelength ($\lambda_s = \lambda_p$) in an isotropic optical fiber with arbitrary polarizations. The propagation of these two waves can be modeled by the following set of nonlinear coupled equations [27]-[28].

$$\frac{\partial u}{\partial t} + v_1 \frac{\partial u}{\partial z} = i\frac{2}{3}\gamma\, v_1 \left[\left(|u|^2 + 2|v|^2\right)u + \left(2|\bar{u}|^2 + 2|\bar{v}|^2\right)u + 2\,\bar{u}\,\bar{v}^*v\right] \tag{1a}$$

$$\frac{\partial v}{\partial t} + v_1 \frac{\partial v}{\partial z} = i\frac{2}{3}\gamma\, v_1 \left[\left(|v|^2 + 2|u|^2\right)v + \left(2|\bar{u}|^2 + 2|\bar{v}|^2\right)v + 2\,\bar{u}^*\bar{v}\,u\right] \tag{1b}$$

$$\frac{\partial \bar{u}}{\partial t} - v_2 \frac{\partial \bar{u}}{\partial z} = i\frac{2}{3}\gamma\, v_2 \left[\left(|\bar{u}|^2 + 2|\bar{v}|^2\right)\bar{u} + \left(2|u|^2 + 2|v|^2\right)\bar{u} + 2\,u\,v^*\bar{v}\right] \tag{1c}$$

$$\frac{\partial \bar{v}}{\partial t} - v_2 \frac{\partial \bar{v}}{\partial z} = i\frac{2}{3}\gamma\, v_2 \left[\left(|\bar{v}|^2 + 2|\bar{u}|^2\right)\bar{v} + \left(2|u|^2 + 2|v|^2\right)\bar{v} + 2\,u^*v\,\bar{u}\right] \tag{1d}$$

where $u$ and $v$ are the left and right circular polarizations of the signal beam whereas $\bar{u}$ and $\bar{v}$ are the circular polarizations of the pump beam, respectively. $\gamma$ is the usual nonlinear Kerr coefficient of the fiber while $v_1$ and $v_2$ represent the group-velocities of the signal and pump waves, respectively. The first four terms on the right-hand side of each equation describe self-phase and cross-phase modulation effects. The last term describes the four wave mixing process responsible for energy exchanges between the two circular polarization components of each wave. As it has been demonstrated in previous works, this exchange of energy can lead to a polarization attraction effect providing that the pump wave is injected into the fiber with a circular polarization. In that case, the signal wave emerges from the fiber with a fixed circular polarization opposed to the pump, independently of its initial input SOP [27]-[28].

As can be seen from equations (1), the nonlinear terms of the right-hand side are basically invariant with respect to a frequency shift of the backward wave, i.e. if $\bar{u}$ and $\bar{v}$ are changed into $\bar{u}\, e^{-i\Delta\omega t + i\Delta\beta z}$ and $\bar{v}\, e^{-i\Delta\omega t + i\Delta\beta z}$, respectively. In fact, as the frequency difference between the two waves increases ($\lambda_s > \lambda_p$), one should essentially modify these equations to include the Raman contribution to the Kerr nonlinearity. The resulting equations can thus be written as [29].

$$\frac{\partial u}{\partial t} + v_1 \frac{\partial u}{\partial z} = i\gamma v_1 \left[ \begin{array}{l} \left(\frac{2}{3}\alpha + \rho\, g_a(0)\right)|u|^2 + \left(\frac{4}{3}\alpha + \rho\, g_a(0) + \rho\, g_b(0)\right)|v|^2 + \left(\frac{4}{3}\alpha + \rho\, g_a(0) + \rho\, g_a(\Omega)\right)|\bar{u}|^2 \\ + \left(\frac{4}{3}\alpha + \rho\, g_a(0) + \rho\, g_b(\Omega)\right)|\bar{v}|^2 \end{array} \right] u$$
$$+ i\gamma v_1 \left[ \left(\frac{4}{3}\alpha + \rho g_a(\Omega) + \rho g_b(0)\right) \right] \bar{u}v\bar{v}^* \qquad (2a)$$

$$\frac{\partial v}{\partial t} + v_1 \frac{\partial v}{\partial z} = i\gamma v_1 \left[ \begin{array}{l} \left(\frac{2}{3}\alpha + \rho\, g_a(0)\right)|v|^2 + \left(\frac{4}{3}\alpha + \rho\, g_a(0) + \rho\, g_b(0)\right)|u|^2 + \left(\frac{4}{3}\alpha + \rho\, g_a(0) + \rho\, g_a(\Omega)\right)|\bar{v}|^2 \\ + \left(\frac{4}{3}\alpha + \rho\, g_a(0) + \rho\, g_b(\Omega)\right)|\bar{u}|^2 \end{array} \right] v$$
$$+ i\gamma v_1 \left[ \left(\frac{4}{3}\alpha + \rho g_a(\Omega) + \rho g_b(0)\right) \right] \bar{v}u\bar{u}^* \qquad (2b)$$

where $\gamma$ is the nonlinear Kerr coefficient, $\rho$ is the fractional Raman contribution to the total nonlinearity, $\alpha = (1-\rho)$ and $\Omega$ is the pump-signal frequency detuning. Similar equations are obtained for the circular components of the pump wave simply by changing z into -z, u into $\bar{u}$ and $v$ into $\bar{v}$. In these equations, $g_a$ and $g_b$ represent the Raman gains associated with co-rotating and counter-rotating circularly polarized waves and are given by :

$$g_a(\Omega) = \int_0^\infty \left( a(t)\exp(i\Omega t) + \frac{b(t)}{2}\exp(i\Omega t) \right) dt \tag{3a}$$

$$g_b(\Omega) = \int_0^\infty b(t)\exp(i\Omega t) dt \tag{3b}$$

where the two Raman response functions a(t) and b(t) can be approximated by [30] :

$$a(t) + b(t) = \frac{\tau_1^2 + \tau_2^2}{\tau_1 \tau_2^2} \exp(-t/\tau_2)\sin(t/\tau_1) \tag{4a}$$

$$b(t) = \frac{2r}{\tau_2}\exp(-t/\tau_2) \tag{4b}$$

In these equations, $\tau_1=32$ fs, $\tau_2=12$ fs and $r$ represents the relative values of the parallel and orthogonal Raman gains in the linear basis components ($r \approx 0.2$).

In a previous study published in 2004 [29], we have demonstrated numerically from these equations and experimentally in the visible domain the possibility of combining Raman amplification and polarization attraction in the same optical fiber. In fact, it has been shown that the counter-propagating polarized pump wave can act both as a polarization attractor and a source of energy for the signal wave. These pioneering experiments were conducted using a short piece of perfectly isotropic fiber with very high peak power nanosecond pulses whose wavelengths were in the visible domain, around 575 nm. In this present paper, our approach is radically novel since we now consider a much longer standard optical fiber, having a low polarization mode dispersion, which allows the device to be compatible with continuous wave, pulse train or telecom processing around 1.55 µm and involving average powers below 1 W. Even though the residual birefringence of such a fiber is very small, it cannot rigorously be considered as perfectly isotropic. Nevertheless, it was shown in a recent work [18]-[22] that such a fiber can be seen as a concatenation of short pieces of isotropic fibers, each acting as a polarization attractor for the signal wave, so that an effective polarization attraction effect can occur at the fiber output thanks to an average phenomenon. This basic approach was confirmed by Sugny *et al.* in refs. [23]-[25]. In fact, one of the most important consequence is that the possibility to use such a long optical fiber results in a significant increase of the effective nonlinear interaction length between the counter-propagating waves, thus involving beam powers compatible with telecom applications. Finally, it is also important to note that the counter-propagating Raman amplification effect is fundamentally not affected by the presence of a weak fiber birefringence. As a consequence, one can expect that such a long low-PMD fiber pumped by an adequate counter-propagating beam could act both as a polarization attractor and a Raman amplifier for the signal wave, thanks to an average process similar to that observed in refs. [18], [21]. In other words, we believe that our system can be considered as a succession of short pieces of isotropic fibers in which the signal is simultaneously repolarized and amplified as described by equations (2). This assumption will be confirmed through the experimental evidences reported in the section below.

## 3. Experimental set-up

Figure 2 illustrates the experimental set-up used to investigate the combined effects of Raman amplification and polarization attraction. At the transmitter side, pulses with full-width at half maximum (FWHM) of 2.5 ps and repetition rate of 10-GHz are first generated using a mode-locked fiber laser (MLFL), operating at a wavelength of 1545 nm. The pulses are then broadened to 45 ps by an optical band pass filter (OBPF) and data encoded by means of an intensity Mach-Zehnder modulator (MZM) driven by a 10-Gb/s $2^7$-1 pseudo-random bit sequence (PRBS). A polarization scrambler is inserted to introduce random polarization fluctuations at a rate of 0.625 kHz. Before injection into the optical fiber, an Erbium doped fiber amplifier (EDFA) was used to reach the suitable average power of 24 dBm (250 mW). The fiber under-test is a 20-km long non-zero dispersion shifted fiber (NZDSF). The dispersion, dispersion slope, nonlinear coefficient and polarization mode dispersion (PMD) are 0.2 ps/nm/km, 0.09 ps/nm$^2$/km, 1.7 (W.km)$^{-1}$ and 0.03 ps/km$^{1/2}$, respectively. The fiber is inserted between two WDM couplers so as to inject and monitor the signal wave. The counter-propagating Raman pump beam, which allows to control the polarization as well as to amplify the signal, was a linearly polarized incoherent wave having a spectral bandwidth of 40-GHz centered around of 1480 nm and with an average power which can be adjusted between 0 and 1 W. At the receiver, the SOP of the signal is analyzed onto the Poincaré sphere by means a commercially available polarization analyzer. Finally, a 30-GHz bandwidth oscilloscope associated with a bit error rate (BER) analyzer allowed to monitor in real-time the intensity profile of the outcoming pulses and transmission quality.

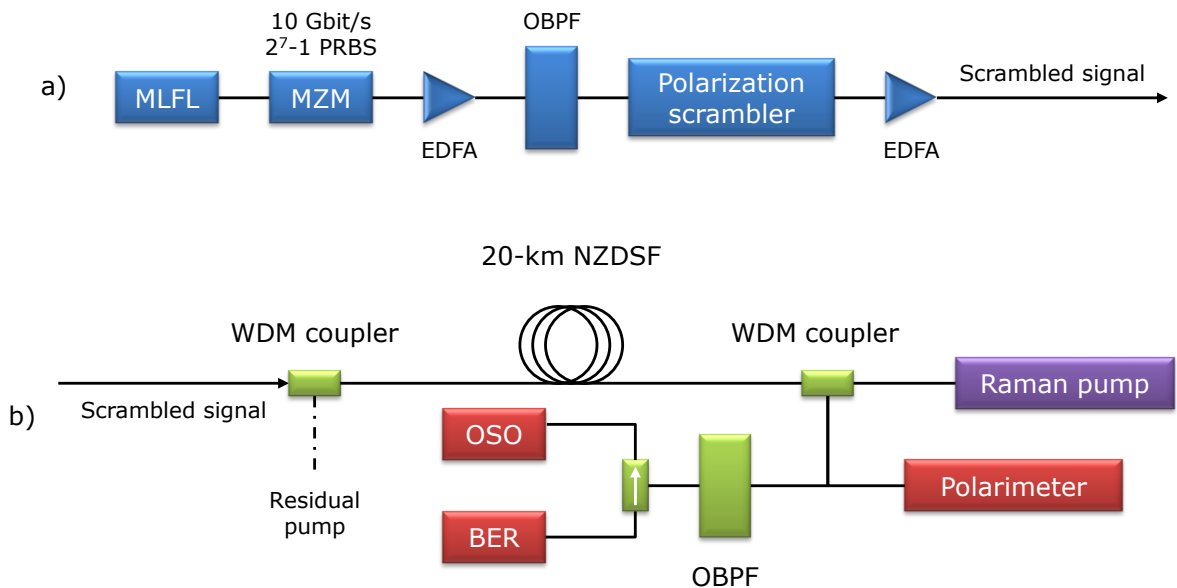

**Fig. 2:** Experimental setup (a) 10 Gb/s RZ transmitter (b) all-optical polarization attraction block. MLFL: mode-locked fiber laser; MZM: Mach-Zehnder Modulator.

## 3. Experimental results: continuous optical signal

Experimental evidences of the combined effects of polarization attraction and Raman amplification were first achieved in the continuous wave (cw) regime at 1548 nm. The state of polarization (SOP) of the signal wave was measured at the output of the fiber as a function of the Raman pump power, for a fixed signal power of 250 mW. Results are shown in Fig. 3 for different pump powers ranging from 0 to 850 mW. In absence of Raman pump (Fig. 3a), and due to the initial polarization scrambling, the different points, representing the signal polarization states, are uniformly distributed onto the Poincaré sphere. As can be seen in Figs. 3b to 3d, when the pump power is gradually increased from 0 to 850 mW, all the signal polarizations asymptotically converges towards the same output state of polarization thanks to the polarization attraction effect (Fig. 3d). Simultaneously, a signal intensity on/off gain of 6.7 dB was measured at the output of the fiber for a pump power of 850 mW, thanks to the counter-propagating Raman amplification effect.

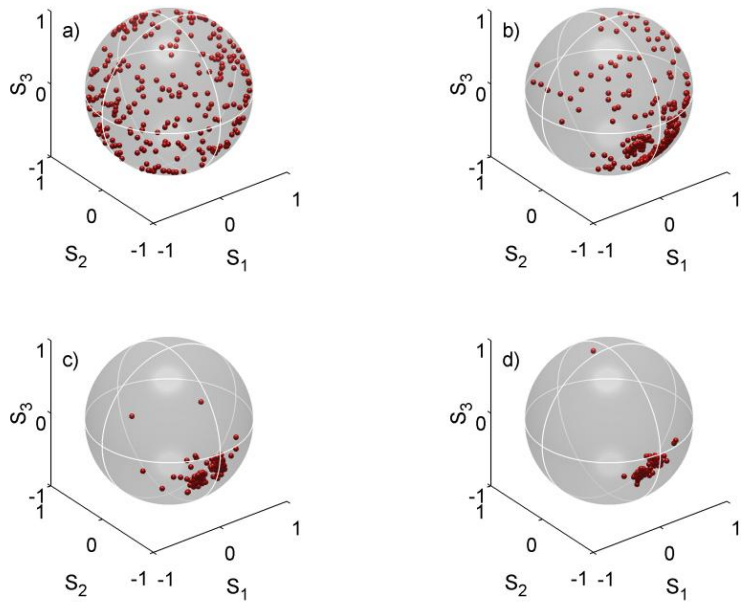

**Fig. 3:** Evolution of the SOP of the output cw signal as a function of the Raman pump power. (a) : 0 mW, (b) : 340 mW, (c) : 680 mW and (d) : 850 mW. The signal power was fixed to 250 mW.

Further information can be obtained by monitoring the output signal wave in the temporal domain thanks to a low-bandwidth oscilloscope. A linear polarizer was inserted just before the photodiode in order to simulate a polarization dependent component, i.e. so that all polarization fluctuations are turned into intensity fluctuations. The oscilloscope was used in persistence mode to superimpose the different waveforms. The results are shown in Fig. 4 for a signal power fixed to 250 mW, without (Fig. 4a) and with (Fig. 4b) the 850-mW counter-propagating Raman pump wave. When the pump wave is turned off (Fig. 4a), the temporal trace shows different levels of intensity, resulting from the different states of polarization induced by the initial polarization scrambling. When the pump wave is turned on (Fig. 4b), the initial power fluctuations induced by the polarization scrambling are significantly reduced, leading to a unique level of intensity at the output of the system and thus demonstrating the ability of our system to reduce polarization fluctuations without any polarization-dependent losses (PDL).

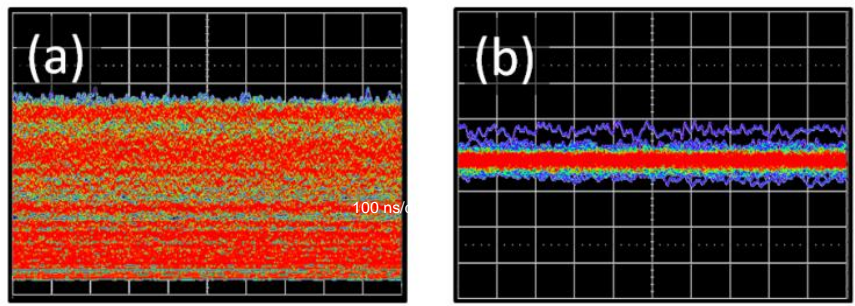

**Fig. 4:** Intensity profile (detected behind a polarizer) of the output CW signal, without (a) and with (b) the 850-mW counter-propagating Raman pump wave. The signal power at the input of the fiber was fixed to 250 mW. Note that the vertical (intensity) scale is not same for (a) and (b) because of Raman amplification occurring in Fig. 4b.

## 4. Experimental results: 10-Gbit/s optical signal

In this section, and in order to underline the practical compatibility of our device with telecommunication applications, we have replaced the initial continuous wave by a 10 Gbit/s PRBS signal wave at 1545 nm. Let us recall that thanks to the counter-propagating configuration of the system, both the polarization attraction and the Raman amplification effects are only sensitive to the average intensity profiles of the signal and pump waves, rather than on the short scale intensity profiles. In fact, the main issue which should be considered when using picosecond pulse trains is the temporal broadening induced by the group-velocity dispersion of the fiber and Kerr nonlinearity. Nevertheless, with the fiber parameters used in our experiment, the excess of temporal broadening observed after propagation in the optical fiber is kept below 10 ps. Figure 5 shows on the Poincaré sphere the SOP of the 10 Gb/s signal at the output of the system as a function of the pump power and for a fixed signal power of 24 dBm (250 mW). As can be seen, all signal polarizations asymptotically converged towards the same output state of polarization when pump power was increased from 0 mW (Fig. 5a), 200 mW (Fig. 5b), 400 mW (Fig. 5c) and finally 850 mW (Fig. 5d). Note that an on/off Raman gain of 6.7 dB was simultaneously achieved when pump power is increased up to 850 mW.

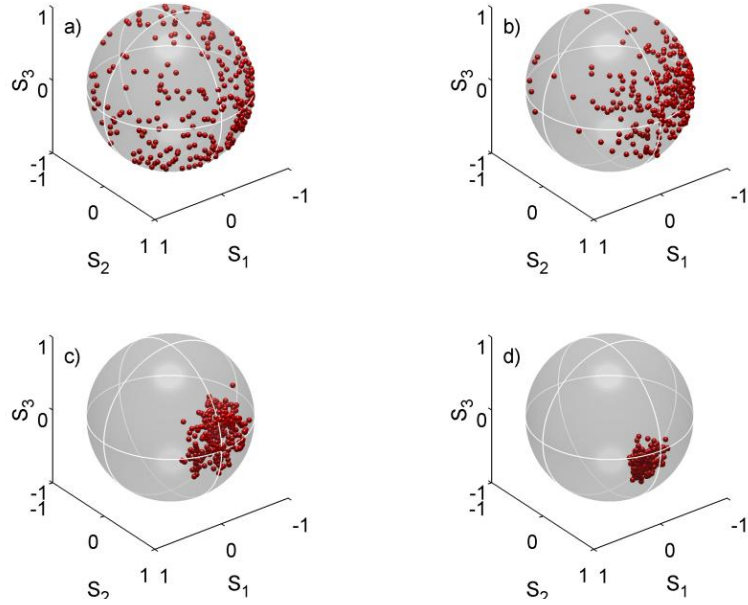

**Fig. 5:** Evolution of the SOP of the 10 Gbit/s signal wave at the output of the fiber as a function of Raman pump power. (a) : 0 mW, (b) : 200 mW, (c) : 400 mW and (d) : 850 mW. The signal power was fixed to 24 dBm. A 6.7 dB on/off gain was achieved for a pump power of 850 mW.

The efficiency of the polarization attraction process can be also evaluated by calculating the degree of polarization (DOP) of the output signal polarization. The DOP is defined as:

$$DOP = \frac{\sqrt{\langle S_1^2 \rangle + \langle S_2^2 \rangle + \langle S_3^2 \rangle}}{S_0},$$

where $S_i$ are the Stokes parameters of the signal and $\langle \ \rangle$ denotes an averaging process over 256 initial conditions and evaluated on a time scale much longer than the duration of the input polarization fluctuations. The DOP permits us to measure the spread of output polarization fluctuations on the Poincaré sphere and thus to quantify the efficiency of the attraction phenomenon. In fact, low values of the DOP indicate that polarization scrambling leads to large temporal fluctuations of the SOP all over the Poincaré sphere. Whereas a DOP value close to unity is associated with a nearly constant and stabilized output SOP. Figure 6 shows the evolution of the output signal DOP as a function of Raman pump power. The DOP of the signal, which has initially low value, close to 0.4, increases strongly when the counter-propagating Raman pump is injected into the fiber so as to reach a value very close to unity for a Raman pump power above 600 mW.

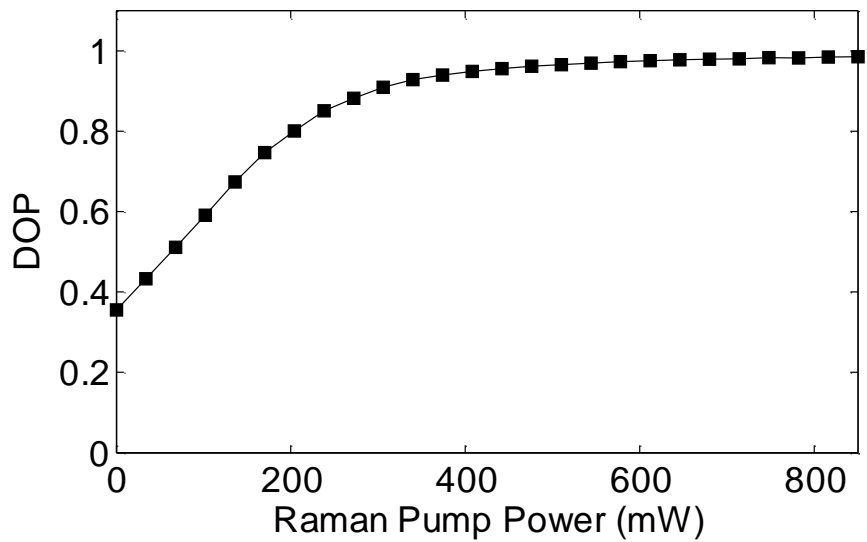

**Fig. 6:** Evolution of the DOP of the 10-Gbit/s signal wave at the output of the fiber as a function of Raman pump power. The signal power was fixed to 250 mW whereas the Raman pump power varies from 0 to 850 W.

The efficiency of the polarization attraction process is more striking when directly monitored in the temporal domain. The 10-Gbit/s eye-diagrams were recorded through a linear polarizer so as to transfer the whole polarization fluctuations directly into the time domain. Resulting eye-diagrams are represented in Figs. 6 in absence (Fig. 6a) and in presence of the counter-propagating Raman pump wave (Fig. 6b). In the pump-free configuration, the initial polarization fluctuations induced by the polarization scrambler are transformed into intensity variations after the polarizer, leading to a complete closure of the eye-diagram (Fig. 6a). On the contrary, when the 850 mW pump beam is injected into the fiber, thanks to the polarization attraction process, all the output pulses emerge from the fiber with the same state of polarization. As a result, the initial intensity fluctuations linked to the polarization variations are greatly reduced, leading to a significant opening of the eye-diagram (Fig. 6b) and recovering of the data. At the same time, the Raman amplification process leads to an intensity gain of the signal beam of nearly 6.7 dB. These results clearly demonstrate the ability of our system to combine simultaneously an all-optical control and stabilization of the polarization state as well as an amplification of a 10 Gbit/s RZ telecom signal.

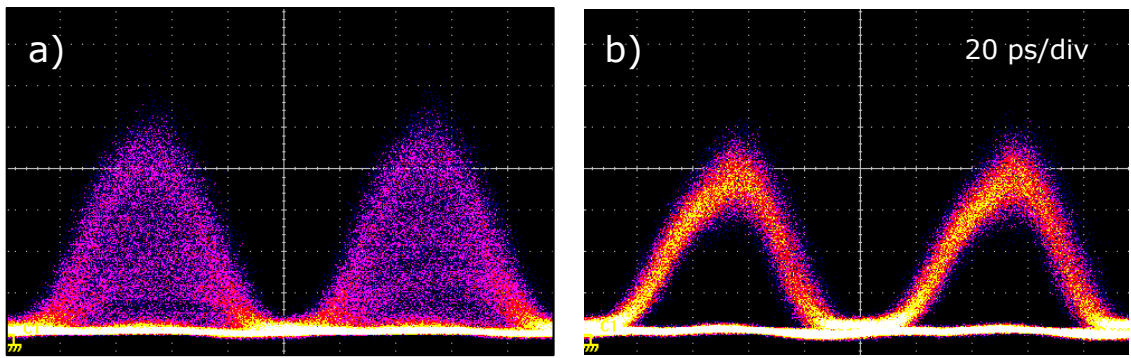

**Fig. 7:** Eye-diagram (detected behind a polarizer) of the output 10 Gbit/s, without (a) and with (b) the counter-propagating Raman pump wave. Signal and pump average power were fixed to 24 dBm and 850 mW, respectively.

Finally, we have also measured the corresponding bit-error-rate (BER) of the 10-Gbit/s signal as a function of the average power incoming on the receiver (Fig. 8). The reference is illustrated by the back-to-back configuration (i.e. at the fiber input) in squares. At the output of the system, when the polarization of the signal is scrambled, corresponding to the eye-diagram of Fig. 6a, the BER is limited to $3.10^{-3}$ (stars). When the 850-mW counter-propagating pump wave is injected into the fiber (circles), the quality of the transmission is greatly improved and low BER penalties were obtained on the receiver.

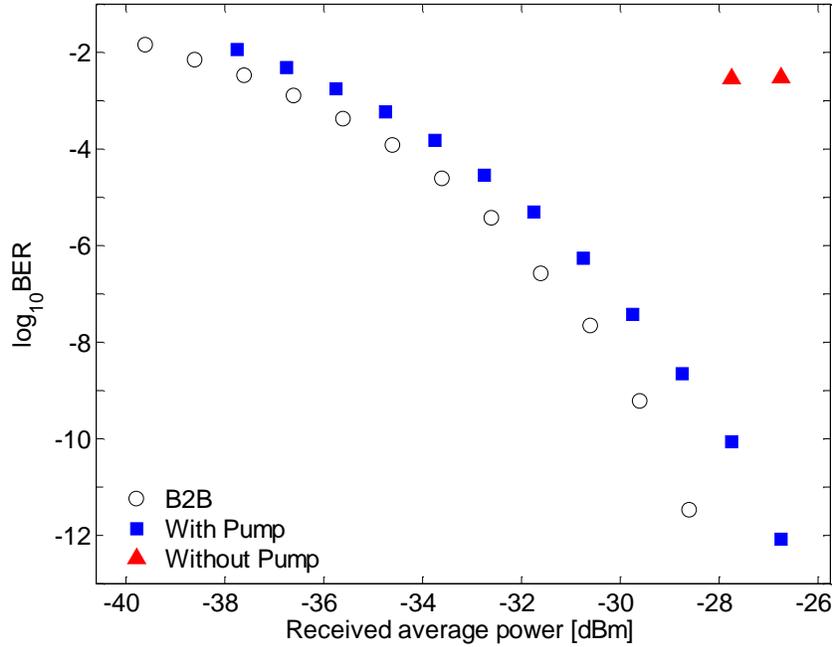

**Fig. 8:** Evolution of the bit error rate as a function of average power in back-to-back configuration (squares); at the output of the system, with polarization scrambling and after a polarizer with (circles) and without (stars) the 850-mW counter-propagating pump beam.

## 5. Conclusion

In this paper, we report an all-optical device which allows combining both polarization attraction and amplification of an optical signal in a unique segment of standard optical fiber. The polarization attraction phenomenon is based on a four-wave mixing process whereas the optical amplification is obtained thanks to stimulated Raman scattering. These two effects were obtained simultaneously in the telecom domain by injecting a polarized counter-propagating pump wave in the optical fiber with a wavelength around 1480 nm. The experimental results were performed with both a continuous signal wave and a 10 Gbit/s PRBS RZ optical signal around 1550 nm.